\documentclass[authoryear,12pt,3p]{jowarticle}
\usepackage{natbib}

\begin{document}
\begin{frontmatter}
\title{Theoretical Equivalence in Physics}
\author{James Owen Weatherall}\ead{weatherj@uci.edu}
\address{Department of Logic and Philosophy of Science \\ University of California, Irvine}
\date{\today}
\begin{abstract}I review the philosophical literature on the question of when two physical theories are equivalent.  This includes a discussion of empirical equivalence, which is often taken to be necessary, and sometimes taken to be sufficient, for theoretical equivalence; and ``interpretational'' equivalence, which is the idea that two theories are equivalent just in case they have the same interpretation.  It also includes a discussion of several formal notions of equivalence that have been considered in the recent philosophical literature, including (generalized) definitional equivalence and categorical equivalence.  The article concludes with a brief discussion of the relationship between equivalence and duality.\end{abstract}
\end{frontmatter}

\section{Introduction}\label{sec:I}

Many of the things we wish to say about the world can be expressed in multiple ways.  We might use different choices of words, for instance, or speak in different languages.  A physical theory may also be expressed in many ways.  For instance, Newton initially published his magnum opus, the \emph{Philosophiae Naturalis Principia Mathematica}, in Latin in 1687; it was later translated into English in 1729 by Andrew Motte and into French in 1749 by \'Emile du Ch\^atelet.  It is hardly tenable to say that each of these translations presents a new physical theory, as opposed to a re-expression of a single theory.

Other cases are more difficult.  Often physicists re-express theoretical claims by changing more than the (natural) language in which they are expressed: they also change the mathematical structures used in formulating the theory.  Such changes in mathematical structure are often accompanied by the introduction of new physical principles, new patterns of inference, and new ways of representing physical systems---even in cases where the new and old formulations of the theory do not differ in their empirical predictions.

For instance, in 1788, the French mathematician Joseph-Louis Lagrange introduced a new formulation of mechanics.  There is a certain sense in which Newtonian and Lagrangian mechanics are empirically equivalent, at least for the systems of interest to Newton and Lagrange.  But whereas on Newton's theory one describes the motions of bodies by specifying the forces acting on those bodies and then using Newton's second law, $F=ma$, to derive the resulting accelerations, on Lagrange's theory one specifies a quantity now known as the ``Lagrangian'', which is a function on possible configurations of a physical system related to the energy associated with that system.  From the Lagrangian one can then derive equations governing the motion of bodies---which turn out to agree, in standard cases, with those you would find from the Newtonian approach.

Did Lagrange introduce a new theory?  Or did he merely re-express, perhaps with some elaboration, the theoretical content of Newtonian physics?  In other words, are Newtonian and Lagrangian mechanics \emph{theoretically equivalent}?  Lagrange's methods certainly allow one to solve more problems, or at least solve some problems more easily, than did Newton's.  In this sense, Lagrange certainly made a novel contribution to classical mechanics.  But whether one should say he proposed a new physical theory depends on what one takes the content of a physical theory to be (in general) and what one takes each of these particular theories to assert about the world.  Many philosophers of physics have argued that the answers to these questions depend, at least in part, on formal or mathematical relationships between the mathematical structures used to represent physical situations in these theories.  Furthermore, these considerations are deeply related, since the formal relationships one takes to capture a suitable notion of theoretical equivalence often reflect a position on theoretical content and interpretation \citep{BarrettThesis,BarrettCM2}.

The question of whether two given theories are equivalent has often arisen in practice.  Heinrich Hertz, for instance, in \emph{The Principles of Mechanics Presented in a New Form}, presents two ``images'' of mechanics that appear to agree in all of their empirical consequences, but which Hertz takes to disagree in the conceptions of the physical world that they suggest \citep{Hertz}; faced with the choice between these images, Hertz suggests a number of extra-theoretical considerations that one might use to distinguish between them.  Similarly, John von Neumann, in \emph{Mathematical Foundations of Quantum Mechanics}, provides an argument that the wave mechanics developed by Schr\"odinger should be taken to be equivalent to the matrix mechanics developed by Heisenberg, Born, Jordan, and others \citep[though see also Dirac \citeyear{Dirac} and Muller \citeyear{MullerMyth1,MullerMyth2}]{vonNeumann}.  More recently, physicists working in high energy theory and string theory have introduced and studied a large number of \emph{dualities}, which are relationships between pairs of theories that appear to differ wildly, but which, by dint of their formal properties, are often taken to provide equivalent descriptions of the same states of affairs \citep{RicklesOld, PolchinskiStudies, deHaroButterfield}.

This article will describe different approaches one might take to the question of when two physical theories are equivalent.  This will include a discussion of empirical equivalence, which is often taken to be necessary, and sometimes taken to be sufficient, for theoretical equivalence; and what I will call ``interpretational'' equivalence, which is the idea that two theories are equivalent just in case they have the same interpretation.  It will also include a discussion of several formal notions of equivalence that have been considered in the recent philosophical literature, including (generalized) definitional equivalence and categorical equivalence.  The article will conclude with a brief discussion of the relationship between equivalence and duality.

\section{Empirical Equivalence}\label{sec:EE}

Suppose that two physical theories both have the resources to describe a given state of affairs, but are such that they differ in the predictions they make for various measurements one might perform.  For instance, both Newtonian gravitational theory and general relativity may be used to describe planetary orbits in our solar system.  They agree, to a high degree of accuracy, concerning the orbits of planets far from the sun, including the Earth.  But the orbits they predict do not agree exactly, and in the case of Mercury, the disagreement is sufficiently significant to be measured using terrestrial telescopes.  (Indeed, the actual orbit of Mercury substantially agrees with that predicted by general relativity, but disagrees with that predicted by Newtonian gravitation.)

In such a case, the two theories in question disagree concerning their empirical predictions: they can be discriminated from one another via observation or experiment.  There is surely no sense in which theories that differ in this way could be said to provide equivalent descriptions of the world.  After all, they appear to provide detectably different descriptions of planetary orbits (among many other situations).  (Of course, there may be situations in which the differences between the theories' predictions in realizable experiments are sufficiently small as to be undetectable, but that is not the point.)  In other words, the theories are \emph{empirically inequivalent}.  Empirical equivalence is often taken as a minimal necessary condition for theoretical equivalence.

Some philosophers of science, mostly working in the first two thirds of the twentieth century, have argued that empirical equivalence is also a \emph{sufficient} condition for theoretical equivalence \citep[eg. ][]{Reichenbach, Grunbaum, Salmon}.  In other words, these philosophers held that what it means for two theories to be equivalent is precisely that they may be applied in the same cases, and when so applied, they always yield the same predictions (or other claims about empirically verifiable matters).

There are several reasons why one might endorse this view.  One is a commitment to some variety of \emph{operationalism} \citep{Bridgman}, which, roughly speaking, is the view that a theory is nothing but a tool for making predictions of a certain sort; philosophers subscribing to operationalism might take it that if two theories agree on their predictions, then they agree on everything that the theory asserts about the world.  Another reason to accept this view would be commitment to \emph{positivism}, which, again roughly, is the view that the only meaningful assertions that one can make are those that are \emph{verifiable}, either through empirical measurement or mathematical proof.  On this view, one might think that the only meaningful assertions that a theory makes are those that are amenable to empirical testing; and so if two theories agree on all of their testable claims, then they agree on all of their meaningful claims.

One does not have to accept a strict verificationist criterion of meaning to endorse a view similar to this last one.  There is a tradition, for instance, originating with \citet{Ramsey} and \citet{Carnap,CarnapBook} and famously articulated and defended by \citet{Lewis,lewis1972psychophysical} (see also \citet{Hempel1,Hempel2}), according to which the meaning of ``theoretical terms'', i.e., terms that are introduced in the course of theorizing, such as ``gravitational field'' or even ``electron'', is to be identified via what is known as a ``Ramsey sentence'' \citep{Psillos}.  Without going into formal detail, a Ramsey sentence is a sentence that asserts that there exists something that has certain properties, which in turn are expressed using terms that we already ``understand''; the theoretical-term-to-be-defined, then, is identified with whatever has those properties already.  Thus, to define ``gravitational field,'' one might say that there exists something that disposes massive bodies to accelerate, depending on their location; and the way in which it so disposes massive bodies is itself determined by the distribution of massive bodies at a time according to a particular equation.  Of course, in this description, we have used other theoretical terms---``massive body'', ``accelerate'', even ``time''---but we presume that those terms are already understood; or else that they, too, may be defined using similar expressions.

Holding something like the Ramsey-Carnap-Lewis view of theoretical terms does not necessarily commit one to the view that empirical equivalence is sufficient for theoretical equivalence, since Ramsey sentences could arguably capture further ``structural'' content of a theory \citep[cf. ][]{MaxwellG1,MaxwellG2,Ketland,Cruse,Melia+Saatsi,Worrall,DewarRamsey}.  But it does raise the question of where Ramsey sentences bottom out, i.e., what sorts of terms or expressions may be taken as sufficiently basic to be used in all theoretical definitions---or even whether such ``basic'' terms exist.   One view, endorsed by some logical positivists and logical empiricists (but not, for instance, by Lewis), is that it is propositions that may be expressed in an ``observation language'', with no theoretical terms at all, that ground theoretical terms.  On this view, even if one were to acknowledge that some expressions that are not empirically verifiable are meaningful, one might nonetheless think that the meanings of theoretical claims ultimately reduce to empirical claims, and thus that if two theories agree on all of their empirical claims, differences between their theoretical claims are matters (only) of definition or convention.

Philosophers of science often cavalierly assert that empirically equivalent alternatives to various theories are possible.  But it is not clear that there is a recipe for finding such theories---or at least, for finding examples that are not trivial notational variants of one another.  Conversely, establishing that two apparently different theories are in fact empirically equivalent or inequivalent is often a subtle matter.  Indeed, even establishing precisely what is meant by ``empirical equivalence'' can be tricky.  For instance, as we indicated above, one would like to say that two theories are empirically equivalent if they can be applied in precisely the same circumstances (and always yield the same predictions).  But in order to assess if this is true, one needs some suitably theory-neutral way of describing possible applications of a theory---i.e., some language for describing physical situations that does not invoke the resources of either theory.  To see the difficulty, consider the question of what predictions Newtonian gravitation makes in the vicinity of a black hole, or what predictions classical electrodynamics and Newtonian mechanics make concerning measurements of an electron's spin.  In such cases, it is hard to see how the theories at issue have the expressive resources necessary to even represent the relevant situations, nevermind the further question of making accurate predictions about them.

It is fair to say in both of these examples that the theories in question are simply empirically inequivalent.  But one needs to be careful, because the mere fact that two theories would give very different theoretical descriptions of a situation does not by itself imply that they are empirically inequivalent, since some systematic translation may be available between different descriptions.  To take an example, the de Boglie-Bohm pilot wave theory (a.k.a. Bohmian mechanics) is widely taken to be empirically equivalent to the standard von Neumann-Dirac formulation of quantum theory \citep{Cushing,Durr+etal}---even though in Bohmian mechanics, particles always have precise and definite positions, and every measurement one makes in the theory is ultimately understood as a measurement of position; whereas in the von Nuemann-Dirac formulation, particles \emph{never} have definite positions, and generically one can measure any number of different quantities.  To establish the empirical equivalence of these theories, one needs to show that measurements of generic observables can always be reconceived as measurements of position.  Such arguments have been given, though they rely on background assumptions about what sorts of measurement apparatuses are, in principle, available.

Reflecting on what the translations needed to establish empirical equivalence involve---in general, the ability to take any situation as described by one theory and identify it with a suitable corresponding description in the other theory, and vice versa---has led some philosophers to argue that empirical equivalence is a much stronger relationship than is usually supposed.  \citet{NortonUnderdetermination}, for instance, has argued that underdetermination of theories by evidence is not a real problem in science, precisely because any two theories that truly could not be distinguished by empirical considerations---that is, theories that were fully empirically equivalent---would necessarily have such a close relationship that one should probably want to say that they were either fully equivalent, or else differed in a way that makes one clearly preferable to the other (say because one posits unnecessary structure or entities). For replies to Norton, see \citet{FrostArnold+Magnus} and \citet{Bradley}.

For further discussion of the interpretational options available when presented with empirically equivalent theories, see \citet{LeBihan+Read}, though be forewarned that they use the term ``duality'' to refer to theories that are empirically equivalent, which is a bit different from the usage here.

\section{Definitional Equivalence and Intertranslatability}\label{sec:DE}

Empirical equivalence is a weak form of equivalence between theories: one might think that two theories could be empirically equivalent, but nonetheless inequivalent in some stronger sense.  For instance, two theories might make the same predictions, but nonetheless differ with regard to what structure they attribute to the world, what sorts of entities exist in the world, or what the laws of nature are.  A well-known example is Bohmian mechanics, which is a quantum theory that arguably makes precisely the same predictions as the standard von Neumann-Dirac formulation of quantum mechanics \citep{Cushing}.  And yet, these theories are rarely taken to be equivalent, since they differ dramatically in their laws, in the sorts of properties objects have, and various other features, such as whether the world is deterministic.  Indeed, many philosophers of physics have argued that the ``standard'' formulation of quantum theory is inconsistent or incoherent, whereas Bohmian mechanics is not \citep{BarrettTheories}.  This suggests that we need a finer grained notion of equivalence.

During the 1970s, \citet{GlymourTETR,GlymourTE} and \citet{Quine} offered proposals for a stronger notion of equivalence between theories.  In both cases, they argued that theories should be taken to be equivalent if they are \emph{inter-translatable}, in the following sense: one should be able to take assertions in one theory and systematically translate them into assertions in the other theory, and vice versa, in a truth-preserving way.  (One might think of this proposal as a generalization of the idea with which we began, that Newton's \emph{Principia} may be translated into English or French without changing the theory.)  Glymour made this proposal precise using the notion of \emph{definitional equivalence}, which is a criterion of equivalence used in first order logic \citep{Montague, Bouvere1, Bouvere2, Kanger, Artigue+etal}.  Quine, too, offered a technical proposal.  But \citet{Barrett+HalvorsonGQ} have recently shown it to have some serious deficiencies and have argued that if one attempts to adjust it so as to capture Quine's original motivation, Quine's proposal collapses into Glymour's.  And so, in what follow we will focus on definitional equivalence and set aside the technical details of Quine's proposal.

Before introducing definitional equivalence in more detail, it is worth emphasizing that both Glymour and Quine, in introducing stronger notions of equivalence between theories, wished to emphasize that theories could be empirically equivalent to one another, but nonetheless theoretically inequivalent, in the sense of making substantially different claims about the world.  For Glymour in particular, this point was made in defense of a variety of scientific realism, according to which physical theories make assertions about the world that extend beyond their direct empirical consequences.  If two theories could be empirically equivalent, but nonetheless inequivalent in some stronger sense, Glymour concluded, then at most one of them could be a true description of the world.  Conversely, Quine was motivated by the question of whether all empirically equivalent theories were necessarily rivals, or if at least some such theories could be seen as equivalent to one another.  (On this latter point, \citet{GlymourTE} also argued that two theories could be empirically equivalent, and yet we could have better confirmation of one than the other.)

We will now turn to defining definitional equivalence.  (For a more precise characterization, see, e.g., \citet{Barrett+HalvorsonME}.)  In the first instance, it is a relationship that can hold between two theories in first order logic.  A first-order theory $T$ consists of two ingredients: (1) a \emph{signature} $\Sigma$, which is a set of symbols, including constant symbols, functions, and predicates (we suppose we have fixed a collection of logical symbols, such as connectives and quantifiers); and (2) a set of \emph{axioms} in that signature, which are formulae (with no free variables) constructed using only the symbols in $\Sigma$, logical connectives, quantifiers, and variables.  The signature can be thought of as a sort of vocabulary: it contains the terms that the theory uses.  And the axioms may be thought of as the basic assertions that the theory makes, expressed using just the vocabulary of the theory.

Now suppose we are given a new symbol, $p$, that is not already in the signature $\Sigma$.  Assume, for the sake of simplicity, that $p$ is a unary predicate, i.e., a symbol such that for any constant $c$, $p(c)$ is a sentence.  An \emph{explicit definition} of $p$ in terms of $\Sigma$ is a sentence (in the signature $\Sigma\cup\{p\}$)
\[
\forall x(p(x) \leftrightarrow \varphi(x)),
\] where $\varphi(x)$ is a formula (with one free variable, $x$) in the signature $\Sigma$.  This sentence asserts that for all $x$, $p(x)$ is true if and only if $\varphi(x)$ holds  Thus $\varphi$ defines $p$ using only the vocabulary available in the theory $T$.  One can define similar sentences for functions and for constant symbols. (The details will not matter, but we remark for completeness that explicit definitions of constants and functions imply certain further sentences, known as admissibility conditions, in $T$.  These conditions reflect constraints on new symbols of a given type,  and so one only considers explicit definitions of new functions and constants for which the admissibility conditions can be proved in $T$.)

By adding new symbols to the signature $\Sigma$, and by adding explicit definitions of those symbols to the axioms of $T$, one can construct a new theory, $T^+$, in a new signature $\Sigma^+$.  Such a new theory is called a \emph{definitional extension} of $T$.  More precisely, a definitional extension of a theory $T$ in signature $\Sigma$ is a theory $T^+$ in signature $\Sigma^+\supset \Sigma$ whose axioms consist of (1) the axioms of $T$ and (2) for each symbol $s\in\Sigma^+$ that does not appear in $\Sigma$, an explicit definition of $s$ in terms of $\Sigma$ (with the property that all of the admissibility conditions for these explicit definitions are satisfied by $T$).  A definitional extension of a theory $T$ is naturally understood to have precisely the same expressive resources as $T$: it says the same things, makes the same sentences (in the signature $\Sigma$) true and false, and so on.  It merely has a larger vocabulary available with which to express the assertions of $T$.

Consider two first order theories, $T_1$ and $T_2$.  Assume, for simplicity, that the signatures of the two theories, $\Sigma_1$ and $\Sigma_2$, are disjoint: they do not both use the term ``electron'', say.  (This assumption is generally benign---though see \citep{Lefever+Szekely}.)  We will say that $T_1$ and $T_2$ are \emph{definitionally equivalent} if there exist definitional extensions $T_1^+$ and $T_2^+$ of $T_1$ and $T_2$, respectively, both in signature $\Sigma^+ = \Sigma_1\cup \Sigma_2$, with the following property: $T_1^+$ and $T_2^+$ are logically equivalent.  Here logical equivalence means that, given any sentence $\varphi$ in $\Sigma^+$, $\varphi$ is provable (so, by the completeness of first order logic, true) in $T_1^+$ if and only if it is provable in $T_2^+$: or, in logical notation, $T_1^+\vdash \varphi$ if and only if $T_2^+\vdash \varphi$.  Observe that it makes sense to ask whether $T_1^+$ and $T_2^+$ are logically equivalent precisely because they have the same signature, and thus they both rule on the truth or falsity of precisely the same set of sentences.  The theories $T_1$ and $T_2$, meanwhile, have different languages, and so one cannot even ask whether they make the same sentences true.

Intuitively, definitional equivalence says that given any sentence $\varphi$ of theory $T_1$, I can, by using explicit definitions of each of the non-logical symbols in $\Sigma_1$ in terms of $\Sigma_2$, translate that sentence into a sentence in the language of $T_2$ that is provable in $T_2$ iff $\varphi$ was provable in $T_1$; and vice versa.  Anything true I can say in one theory can be said, equally well and with the same truth conditions, in the other theory.  In this way, it captures Glymour's (and, to an extent, Quine's) suggestion that two theories should be equivalent if (and only if) they are inter-translatable.

Definitional equivalence certainly captures an interesting and important sense in which two theories may be equivalent, at least in first order logic.  But is has some problems.  The most immediate problem is that definitional equivalence applies to first order theories, while we are interested in equivalence between \emph{physical} theories, which are rarely expressed in first order logic.  Worse, many of the mathematical tools that we regularly use in physics, such as topology, apparently do not have first order axiomatizations.  And even if they did, it is not clear that we could capture the full range of theoretical practice in physics in the framework of first order logic.  If we do not have, once and for all, a ``language'' or ``axioms'' for each of our physical theories, how could we hope to apply definitional equivalence in practice?

\citet{GlymourTE} recognized this concern.  But, he argued, definitional equivalence could still be useful in physics. He illustrated this claim with an example from Newtonian gravitation.  First, Glymour pointed out that while definitional equivalence is a syntactic notion, it has a semantic counterpart, using the theory of models of first order logic \citep{Hodges, hodges1997shorter}.  Roughly speaking, a \emph{model} of a first order theory $T$ is an ordered collection of sets, consisting of a domain of quantification (i.e., a set of objects about which the theory is interpreted to make assertions) and subsets of that domain (or subsets of products of that domain with itself), corresponding to the extensions of each of the various predicates, functions, and relations in the theory's signature, $\Sigma$.  A model may be thought of as a structure, along with an interpretation of the sentences of $T$ as assertions about that structure, such that those assertions are all true.

How can we express definitional equivalence using models?  To answer this, we need two facts.  First, observe that another, equivalent characterization of logical equivalence of two theories in the same signature is that they have (precisely) the same models.  Second, observe that, given any theory $T$, any model $M$ of $T$, and any definitional extension $T^+$ of $T$, there exists a unique \emph{definitional expansion} of $M$, $M^+$, which is a model of $T^+$ \citep[p.53]{hodges1997shorter}.  Thus, I can take the models of $T$---the structures of which $T$ is true---and uniquely expand them to yield new structures of which the extended theory $T^+$ is true.  Putting these pieces together, we can conclude that theories $T_1$ and $T_2$ are definitionally equivalent only if for every model $M_1$ of $T_1$, there is a unique definitional expansion $M^+_1$ of $M_1$ to a $\Sigma_1\cup\Sigma_2$ structure,  whose ``reduct'', i.e., restriction to $\Sigma_2$, is a model of $T_2$, and vice versa.

This semantic version of definitional equivalence does not immediately solve the first problem above, because it still concerns first order logic.  But it is suggestive, because one does, in physics, often have ``models'' of a physical theory---and these models are generally mathematical structures that ``realize'' the assertions of the theory. The models of general relativity, for instance, are smooth, four dimensional manifolds (satisfying some further conditions) endowed with smooth, Lorentz-signature metrics.  The models of non-relativistic quantum mechanics are Hilbert spaces, along with a suitable subalgebra of the bounded operators on that Hilbert space.  And so on. (One has to be careful here, because a ``model'' in physics is not generally a $\Sigma$-structure for any first order theory; indeed, scientists and philosophers of science use the term ``model'' in an enormous variety of ways \citep{Downes,Weisberg,OConnor+Weatherall}.)

In light of these considerations, Glymour proposed a ``liberal'' notion of definitional equivalence---one inspired by the ideas of first order logic, but applicable more broadly.  He did not precisely define this notion of equivalence, but he did argue that it should imply the following: two theories are equivalent only if, given any model of one of the theories, one can uniquely transform it into a model of the other theory; and vice versa.  Thus we find, if not a sufficient condition, at least a necessary condition for theoretical equivalence that can be applied in real cases.

\citet{GlymourTE} illustrated the point with an influential example.  Newtonian gravitation is standardly formulated as a theory on in which bodies accelerate in the presence of a gravitational field, which in turn depends on the distribution of matter.  But there is another formulation of the theory, of interest to philosophers because it bears some qualitative similarities to general relativity, on which gravitation is ``geometrized''.  In this theory, there is no gravitational field and bodies do not accelerate due to gravitation; instead, they follow geodesics in curved spacetime, with curvature proportional to the distribution of mass \citep[cf. e.g.,] [Ch. 4]{MalamentGR}.  One might then ask: are these two theories equivalent to one another?  There is a classic result due to \citet{Trautman} that bears on this.  It says: given any model of standard Newtonian gravitation, there exists a unique model of the geometrized theory with the same mass density in which bodies follow the same trajectories; and conversely, given any model of the geometrized theory (satisfying certain conditions), there exists a model of the standard theory that, likewise, agrees on mass density and trajectories.  Thus, we have a precise sense in which the theories are empirically equivalent, and we can even translate between them in a certain sense.  But, Glymour argues, they are \emph{not} equivalent according to the criterion of definitional equivalence.  The reason is that while the translation from the standard theory to the geometrized theory is unique, it is many-to-one, and so the translation in the opposite direction is not unique.  Applying definitional equivalence in this sort of way may not be dispositive, but it certainly seems probative.

In this example, Glymour uses definitional equivalence as a necessary condition.  But does definitional equivalence also provide us with sufficient conditions for equivalence?  There is reason to think the answer is ``no'', at least so far.  As described, definitional equivalence is a purely formal relation between theories.  Nothing in the notion of ``explicit definition'' or the ensuring characterization of definitional equivalence requires that translations between theories preserve any prior ``meaning'' of the terms or sentences of either theory.  But, as \citet{Sklar} pointed out in an influential response to Glymour, physical theories have (physical) interpretations; presumably two theories are equivalent only if their interpreted claims about the physical world are in some good sense the same.  Consider, for instance, the mathematical theory of Brownian motion as applied to a particle of pollen suspended in a fluid, and the same mathematical theory as applied to stock market prices \citep[cf. ][p. 279]{BvF}.  There is surely some sense in which these theories are ``intertranslatable'': every time the word ``position'' appears in the first theory, one can substitute ``price'' in the second, and so on.  And yet they are surely not equivalent theories since they are talking about different subject matter.

We will return to the issue of ``interpretational equivalence'' in section \ref{sec:IE}, but at least in the first instance, it seems that if definitional equivalence is to be a satisfactory criterion of equivalence between physical theories, it must at very least be supplemented with a requirement that equivalent theories be empirically equivalent, and moreover, that the translations between the theories that realize the definitional equivalence be suitably compatible with this empirical equivalence.  It is not clear how to make this requirement precise, and in any case, as we saw above, empirical equivalence is itself a difficult notion.  But it is also true that, as in the case of Newtonian gravitation just mentioned, in some cases theories are empirically equivalent, and moreover, that one can require definitional equivalence as a further, strictly stronger condition.

We conclude this section by remarking on another problem with definitional equivalence, which is that it is arguably \emph{too strong}, even in the case of first order theories.  For instance, as \citet{Barrett+HalvorsonME} note, definitional equivalence cannot capture cases in which two theories that use (multiple) different \emph{sorts}, i.e., different classes of entity to which different predicates apply, might be equivalent. \citep[See also][for a concrete example in which the difference matters.]{barrett2017geometry}  Instead, they propose a weaker notion of equivalence that they call \emph{Morita equivalence}, but which others have called \emph{generalized definitional equivalence} \citep{andreka+etal2002, Madarasz, Andreka+NemetiComparing}.
Similarly, \citet{Andreka+NemetiComparing} and \citet{Lefever+Szekely} argue that definitional equivalence needs to be generalized to accommodate theories with non-disjoint signatures.

These considerations lead to a number of related notions of equivalence, all motivated by similar considerations as definitional equivalence, that can help decide cases of theories in first order logic (as, for instance, in \citet{Lefever+SzekelyComparing}).  In general, however, the difficulties described above concerning how to apply these notions to cases in which a first order formulation is not available remain open.

\section{Categorical Equivalence}\label{sec:CE}

As noted at the end of the previous section, there are reasons to think that definitional equivalence is either inadequate or of limited use in practice (or both).  Motivated by these and related concerns, \citet{Halvorson} and \citet{WeatherallTE} have proposed a different criterion of equivalence between physical theories, using methods from category theory; this proposal has subsequently been pursued and developed by numerous authors \citep{HalvorsonOxford,HalvorsonLandryVolume,Rosenstock+etal, Nguyen+etal, Dewar+Eva, BarrettCM1,BarrettSTS, BarrettCM2, WeatherallLandry, WeatherallUG}.\footnote{The development of the subject is difficult to establish based on the published literature, since a large number of papers either appeared online or were published at approximately the same time, even though they were produced sequentially over the course of about five years.  Briefly, Halvorson first publicly discussed the ideas that led to \citep{Halvorson} in a February 2011 lecture in Irvine; \citep{WeatherallTE} was first drafted the following summer (in close conversation with Halvorson while on a visit to Princeton) and distributed to the Southern California Philosophy of Physics Group and others in Fall 2011.  It went on the Pittsburgh Philosophy of Science Archive in 2014 and was accepted for publication in early 2015, but was not published until 2016.  In the meantime, a number of papers that drew on, extended, and criticized the ideas of \citet{Halvorson} and \citet{WeatherallTE}---such as \citet{Rosenstock+etal}, \citet{Rosenstock+Weatherall}, \citet{HalvorsonOxford,HalvorsonCategories}, \citet{Barrett+HalvorsonGQ, Barrett+HalvorsonME}---were drafted, circulated, and published, with many appearing online in 2016.}

Categorical equivalence has a number of virtues, including some that definitional equivalence lacks.  For instance, it can be readily applied to real cases, and it appears to render intuitively plausible verdicts in a number of cases of real interest.  Moreover, it captures senses of equivalence that have been implicitly invoked in earlier philosophical literature \citep{Ryno, Rosenstock+etal, WeatherallUG}, and related notions of equivalence are often used in mathematical physics \citep[eg. ][]{Schreiber+Waldorff,Schreiber} and, recently, in the foundations of mathematics \citep[eg. ][]{hottbook,Visser}.

Before introducing categorical equivalence, we remark on the motivation for the proposal.  \citet{Halvorson} begins with a critique of the so-called ``semantic view of theories'', according to which a theory should be identified with a collection of models, on the grounds that it delivers the wrong verdicts on equivalence, as seen in a series of examples.  (Recall the discussion above concerning the ambiguity in the meaning of the term ``model'' here.)  At the end of the article, he sketches an alternative proposal: to adequately capture the structure of a theory, he suggests, one needs to consider \emph{structured} sets of models, i.e., collections of models with further information about the relationship between those models.  Invoking recent work in categorical logic \citep{Makkai,AwodeyForssell}, he suggests that categories of models may be natural candidates to represent theories.  (In reply, \citet{GlymourHalvorson} argued that in fact \emph{definitional} equivalence provides a suitably semantic characterization of equivalence (recall section \ref{sec:DE}); to which \citet{HalvorsonGlymour} replied that definitional equivalence, even formulated using tools from model theory, invokes the signature of the theory in a way that is in tension with the rhetoric of the semantic view's defenders.  For our purposes, this is a side issue; but see \citet{lutz2014s,lutz2017syntax,BvF,hudetz2017semantic,Walsh+Button} for further discussion.)

\citet{WeatherallTE} comes to categorical equivalence by reflecting on examples.  Recall that Glymour applied definitional equivalence to study the relationship between ``standard'' and geometrized Newtonian gravitation, concluding that these theories cannot be equivalent.  Weatherall first shows that if one applies Glymour's argument to another pair of theories---classical electromagnetism in a ``gauge dependent'' formulation, in terms of 4-potentials; and classical electromagnetism in a ``gauge free'' formulation, in terms of an electromagnetic field strength (Faraday) tensor---one likewise concludes that the theories are inequivalent.  And the reason is the same: there is a many-one relationship between 4-potentials and electromagnetic fields.

The trouble with this conclusion is that physicists are well-aware of the asymmetry between potentials and electromagnetic fields, and yet the two formulations of electromagnetism are used interchangeably. As Weatherall argues, the reason is simple: physicists recognize additional relationships between the models of the theories.  In particular, 4-potentials associated with the same electromagnetic field are understood to be equivalent, in the sense that they can be used interchangeably to represent the same physical situations.  Such 4-potentials are said to be related by \emph{gauge transformations}.  (Note that there are two senses of equivalence under discussion, now: one is a relationship between \emph{theories}, or formulations of theories, while the other is between models of a single theory.)

Weatherall then proposes adjusting Glymour's criterion so that two theories are equivalent if the models of one can be uniquely transformed into models of the other, and vice versa, in a way that takes you back to the model with which you began, \emph{up to model equivalence}.  Applying this moral to Newtonian gravitation, then, Weatherall argues that whether standard and geometrized Newtonian gravitation are equivalent depends on a prior choice of whether models of the standard theory associated to a single model of the geometrized theory should be taken as equivalent.  If one concludes that they are---which one might have independent motivation for doing \citep{NortonCosmology,MalamentNorton,NortonAcceleration,WallaceNewtCosm}---then the two theories should be taken to be equivalent; if not, then Glymour's original argument stands.

This line of thought has just been rehearsed without invoking any category theory.  So where does categorical equivalence come in?  It turns out that categorical equivalence is a way to make the idea of ``unique transformation up to equivalence'' precise.

A \emph{category} consists of two sort of data: a collection of \emph{objects} and, for each ordered pair of objects $A$ and $B$, a collection of \emph{arrows} between them, denoted $hom(A,B)$. These are required to satisfy the following criteria: given arrows $f$ and $g$, such that $f$ terminates at the object at which $g$ originates, then there exists a unique arrow, $g\circ f$, called the \emph{composition} of $g$ with $f$, originating at the same object as $f$ and terminating at the same object as $g$; composition of arrows is associative, in the sense that, given suitably composable arrows $f$, $g$, and $h$, we have $h\circ(g\circ f)=(h\circ g)\circ f$; and finally, for every object $A$, there exists a unique arrow \emph{identity arrow} $1_A$, originating and terminating at $A$, with the properties that for any $f$ originating at $A$, $f\circ 1_A = f$, and for any arrow $g$ terminating at $A$, $1_A\circ g= g$.  \citep[For further details on category theory, see][]{MacLane,Awodey,Leinster}

On a first pass, it is helpful to think about so-called ``concrete'' categories, to get a sense of how they work.  Consider, for instance, the category \textbf{Set}, which has, as objects, sets, and as arrows, functions between sets; or the category \textbf{Group}, which has, as objects, groups, and as arrows, group homomorphisms.  (Note that there is a problem, here, related to ``size'', since there is no set of all sets.  We set this aside for present purposes.)

A \emph{functor} is a map between categories that takes objects to objects, arrows to arrows, and which preserves domains, codomains, composition, and identity.  Functors can have various properties, analogous to the way in which functions (between sets, say) may be ``injective'' or ``surjective''.  Fix a functor $F:C\rightarrow D$.  We will say that $F$ is \emph{full} if, given any two objects $A$ and $B$ of $C$, the map that $F$ induces between $hom(A,B)$ and $hom(F(A),F(B))$ is surjective.  It is \emph{faithful} if that map is injective.  And it is \emph{essentially surjective} if, for every object $X$ of $D$, there is an object $A$ of $C$ such that $F(A)$ is isomorphic to $X$, where by ``isomorphism'', here, we mean that there exist arrows $f$ and $g$, from $F(A)$ to $X$ and vice versa, respectively, such that $f\circ g=1_X$ and $g\circ f=1_{F(A)}$.  Now suppose we are given two categories, $C$ and $D$.  We say that $C$ and $D$ are \emph{equivalent} if there exists a functor $F:C\rightarrow D$ that is full, faithful, and essentially surjective.  If this holds, then there exists a functor $G:D\rightarrow C$ that is also full, faithful, and essentially surjective that is ``almost inverse'' to $F$, in the sense that the composition $G \circ F:C\rightarrow C$ takes objects of $C$ to isomorphic objects of $C$, and likewise for $F\circ G:D\rightarrow D$.

What does this relationship capture?  For one, it says that the two categories are ``almost'' isomorphic, in the sense that the objects of $C$ and $D$ can be uniquely identified with one another ``up to isomorphism'', in a way that preserves all of the relations between objects encoded in the arrows of the categories.  This sort of ``almost isomorphism'' of categories turns out to be enormously fruitful in mathematics. Indeed, many deep theorems relating disparate areas of mathematics turn out to be expressible as assertions of equivalence (or other functorial relations) between categories.

Intuitively, the idea of categorical equivalence is close to the ``up to isomorphism'' relationship described between the formulations of electromagnetism discussed above.  And indeed, this idea can be made precise, by defining two categories: one of which has as objects electromagnetic field tensors on Minkowski spacetime, and has as arrows ``isomorphisms'' of that structure; and another has as objects 4-potentials, and has as arrows ``isomorphisms'' of that structure, including gauge transformations.  \citep[See ][for further details.]{WeatherallTE,WeatherallUG,Nguyen+etal}  One then finds that these categories are equivalent, with the equivalence realized by functors whose action on objects is given by the relationship described above.  If, on the other hand, one did not include the gauge transformations, the resulting categories would not be equivalent.  Similarly for Newtonian gravitation.

Observe that the functors just described may also be said to preserve ``empirical structure'', i.e., they take models of one formulation to models of the other that have the same observational significance.  One can spell this out in several ways, but they all turn on the same idea: ultimately, the empirical significance of a model of electromagnetism is entirely encoded in the electromagnetic field strength.  Since the functor described preserves the electromagnetic field strength associated with models in each formulation, the functor must preserve empirical structure, or ``predictions''.  Presumably, this is a condition we should demand on any functor that is a candidate for realizing an ``equivalence'' between theories, for the same reasons we considered above in connection with definitional equivalence.

We can extract from this discussion a candidate criterion of theoretical equivalence.  Suppose we are given two theories, each represented as categories whose objects are the models of the theories and whose arrows preserve, in some suitable sense, the structure of the models.  We say that the two theories are (categorically) equivalent if there exists an equivalence of these categories that preserves empirical structure in the sense just sketched \citep{WeatherallTE}.

What considerations recommend this criterion?  Many of our physical theories lend themselves to characterizations as collections of certain mathematical structures.  If categorical equivalence is a fruitful notion of equivalence between mathematical theories, then it can presumably be used to capture a sense in which the mathematical structures used in those theories are equivalent (qua mathematics); and if, in addition, the physical theories are empirically equivalent, in the sense described above, one captures a sense in which categorically equivalent theories use equivalent mathematics to capture the same empirical regularities.  (Of course, this argument depends on a prior acceptance of categorical equivalence as fruitful in mathematics, which has not been defended here.)  Moreover, categorical equivalence is readily applicable to many cases \citep{WeatherallLandry}: in addition to the cases already discussed, these methods have clarified the sense in which Hamiltonian mechanics is equivalent to Lagrangian mechanics \citep{BarrettCM2} and general relativity is equivalent (actually, dual) to the theory of Einstein algebras \citep{Rosenstock+etal}.  Finally, categorical equivalence offers a fruitful guide to the ways in which two otherwise similar theories \emph{fail} to be equivalent.  In particular a given functor may fail to be full, to be faithful, or to be essentially surjective (or more than one of these conditions may fail); such functors are sometimes said to be \emph{forgetful} \citep{Baez+etal}.  Studying the properties of such functors can allow one to say what is ``forgotten'' when moving from one theory to the other \citep{WeatherallUG,Rosenstock+Weatherall,Nguyen+etal,Bradley+etal}.

But there are also reasons to be cautious about this criterion of equivalence.  One problem is how to choose the right categories in the first place.  Arguably, many physical theories can be expressed by describing a collection of models.  But to construct a category, one needs to provide additional information: arrows between these models.  And as we have already seen, there are often multiple choices available.   How do we know we have constructed the correct category of models for a given theory?  One reason why this problem may not be devastating is that, in the cases we have considered, different choices reflect different ways of understanding the theory itself, in a way that may draw attention to salient interpretational ambiguities.  We saw this already in the cases of electromagnetism and Newtonian gravitation; \citet{BarrettCM2} has made the point particularly clearly in the context of Lagrangian and Hamiltonian mechanics---two theories whose relations have been a matter of some dispute in the recent literature \citep{North,Curiel,BarrettCM1}.

There is another version of this worry, however, that may have more teeth.  Categorical equivalence makes sense only if we represent a theory by a category that adequately represents the structure of that theory.  Under what circumstances can we be confident that we have done so?  For instance, given a category of models, and no further information, can one reconstruct a theory?   In particular, we usually think of the ``internal'' structure of models of a theory as representing physical situations---for instance, in general relativity it is the points of a manifold that represent events in space and time.  But it is not always possible to reconstruct this internal structure from the categorical structure---that is, from the arrows between models.  This worry has been sharpened and emphasized by \citet{Hudetz}, who has argued that simply identifying a functor between theories that is full, faithful, and essentially surjective is sometimes not sufficient to establish that the theories are equivalent, even if the functor preserves empirical equivalence.  The reason is that such a functor may not take models to other models with suitably related internal structure.  To see this worry, consider an extreme case: define a category of models of some physical theory, and then take another category whose objects are ``structureless points'' and whose arrows are chosen precisely so as to make the two categories equivalent.  The latter category arguably does not represent a physical theory at all, much less one equivalent to what one began with.

To avoid this trivialization worry, Hudetz proposes that the functor needs to be what he calls a ``reconstruction functor'', which means that the model-to-model mapping determined by the functor must take models to models that can be ``reconstructed'' from the model with which one began.  The notion of ``reconstruction'' offered here is very similar to that used in section \ref{sec:DE}, in ``defining'' models of one theory from models of another in the context of definitional equivalence or generalized definitional equivalence.

Hudetz's proposal may be thought of as a hybrid of (generalized) definitional equivalence and categorical equivalence, and it arguably has the advantages of both.  On the other hand, it also inherits disadvantages of both.  Moreover, it is not clear that the trivialization worry is real, at least if one has chosen the categories one begins with judiciously: for a sufficiently rich category, one may be able to ``reconstruct'' the internal structure of any object in a category, up to isomorphism, from the arrows of the category.  (Indeed, one can do precisely this in, for instance, the category of sets and functions; see also \citet{barrett2017symmetries} for a discussion of how much structure is captured by the isomorphisms between models of a first order theory.)  From this perspective, the ``structureless points'' of the example above are not structureless after all, and the further restriction to reconstruction functors is unnecessary or, in other words, automatic.  (For further discussion of worries along these lines, see \citet{WeatherallWNCE}.)

There are many open questions, both mathematical and methodological, related to these last two arguments.  But they both point at the same worry: categorical equivalence may be too weak.  (Indeed, \citet{Barrett+HalvorsonME} explicitly show that it is strictly weaker than generalized definitional equivalence in the first order case.)  One would like to have better control over when categorical equivalence yields reliable guidance, and when it does not.  But in the meantime, there is another attitude one can adopt, defended by Sarita Rosenstock.  On her view, categorical equivalence is not a formal criterion of equivalence that one can use, with no background understanding of the theories involves, to individuate theories.  Instead, it is a heuristic for evaluating proposed relationships between theories of prior interest.  In other words, the right question to ask is: given two theories and some proposed relationship between them, can one capture that relationship as a certain functor between categories associated with the theories, and if so, is that functor full, faithful, and essentially surjective?  (One could also extend this, and ask: is the functor a reconstruction functor in Hudetz's sense, above?)  If so, that suggests that the relationship may realize a sense in which the theories are equivalent; if not, then one can use the properties of the functor to clarify what is ``forgotten'' as one moves between the theories.

\section{Interpretational Equivalence}\label{sec:IE}

(Generalized) definitional and categorical equivalence rely on formal relationships that might obtain between pairs of scientific theories, expressed in a particular way (e.g., as a first order theory).  But some authors have recently argued that formal criteria of equivalence cannot succeed.  In particular, \citet{Coffey} has argued that whether two theories are equivalent---or perhaps better, whether two theories are different presentations of a single underlying theory---is a question of how we interpret the theories: that is, do these theories represent the same physical ontology, possibly structured in the same ways, governed by the same laws, etc.?  (For a closely related view, see \citet{Maudlin}.)  This is not a question about formulations of a theory, so much as a question about our intentions and interpretations; as such, the question cannot be settled by formal criteria.  Coffey argues that recognizing the role of interpretation in judgments of equivalence and inequivalence can explain why there are disagreements about particular cases of putative equivalence---including the examples of electromagnetism and Newtonian gravitation discussed above---and why in some cases there are asymmetries in these judgments, such that, for instance, one may need to ``fix'' one formulation of a theory, for instance by introducing gauge transformations.  In such cases, interpretation of the formalism is surely playing an essential role in guiding our understanding of what ``parts'' of a theory to take seriously for purposes of judging their equivalence.  \citep[See also ][ discussed in section \ref{sec:D}]{Butterfield,LeBihan+Read}

Along similar lines---but beginning with the literature on scientific representation---\citet{NguyenTE} has recently argued that whether two theories are equivalent should be understood as a question about whether they permit one to model the same ``target systems'', and, if they do, whether the models they offer warrant the same claims or inferences about those target systems.  He, too, suggests that ``purely formal'' accounts of theoretical equivalence cannot be sufficient, because they cannot capture the role that intention and interpretation play in the semantics of scientific theories and models.  His alternative proposal is that two theories are equivalent precisely when they permit the same claims and inferences under the same circumstances.

Do these sorts of challenges pose threats to the formal criteria discussed above?  Yes and no.  On the one hand, there is clearly something right about Coffey and Nguyen's arguments, and the earlier arguments of \citet{Sklar}, that any adequate account of theoretical equivalence will need to respect the semantics of physical theories, including what we interpret them to be saying about the world.  If, when the dust settles, one wishes to say that two theories make substantially different claims or warrant different inferences in a given situation, one surely would not wish to say they are equivalent.  On the other hand, it is not clear how much tension there is between these views and the criteria already discussed.  In particular, it bears emphasizing that neither definitional equivalence nor categorical equivalence is a \emph{purely} formal criterion, at least as applied in the recent literature, insofar as empirical equivalence is also taken as a necessary condition for both, and, as we noted above, specifying the empirical content of a theory is itself a difficult task that connects to (though does not exhaust) issues in the semantics and interpretation of scientific theories.

Where perhaps there \emph{is} disagreement concerns how to undertake the project of interpreting physical theories in the first place, and whether questions of (formal) equivalence and inequivalence have a role to play in those discussions.  If one has the view that interpretation is ``easy'', in the sense that one can read off from either the formalism of a theory, or perhaps the formalism plus sociological data about what inferences are drawn from that formalism, what the theory ``says'' about the world, then presumably there is no need for formal notions of equivalence of theories: one can simply ask, of two theories, whether they ``say'' the same thing.  Interpretational equivalence is the end of the story.

If, on the other hand, one thinks that there are challenges in extracting from a theory's formalism the theoretical claims that it makes, or if one thinks that it is possible to say or represent the same things in apparently different ways, then even if one endorses the idea that theories are equivalent just in case they have the same interpretation, one might think that formal criteria for equivalence can and do play an important role in establishing what the interpretational options are.  (Recall that in the case of categorical equivalence, at least, representing a theory by a category itself raised interpretational questions that may otherwise have been obscure; formulating a first-order formalization of a theory will generally do the same.) On this point, \citet{BarrettThesis,BarrettCM2} has argued that there is a close relationship between one's attitudes concerning what it means to interpret a theory and when two theories are equivalent, and suggests that exploring different criteria for equivalence is a proxy for exploring different strategies for interpreting physical theories in the first place. Finally, one might argue that interpretational or representational equivalence merely defers the difficult issues, since now one needs to establish when two interpretations are equivalent---and presumably interpretations, like theories, may be expressed in multiple ways, for example by using different languages.

\section{Duality}\label{sec:D}

The discussion above has concerned issues of theoretical equivalence as they have played out in philosophy of science, particularly over the last four decades.  But over the same period, a parallel discussion of the meaning and significance of ``distinct but (somehow) equivalent'' theories has occurred within physics.  Physicists often say that such theories are ``dual'', or that pairs of such theories exhibit (or are) a ``duality''.  Dual descriptions of the same physical situations have become particularly important in the context of string theory, which is an ambitious program to unify gravitational physics with the Standard Model of particle physics \citep{PolchinskiStudies}.  Still more recently, philosophers of physics have attempted to understand the character and interpretational significance of these dualities \citep{Matsubara, Read, RicklesOld,Rickles,Read+MN,deHaroConceptual,deHaroDEG,deHaroTehButterfield1,deHaroTehButterfield2,deHaroButterfield,Butterfield,LeBihan+Read}.  We will not give a complete review of the issues related to dualities here; instead, we will focus on how the literature on dualities relates to the issues of equivalence already introduced.  (For a recent review of other philosophical issues related to dualities, see \citet{LeBihan+Read}, though as noted above, they use the term ``duality'' to refer to any empirically equivalent theories, which is somewhat different from the usage adopted here.)

The contemporary notion of duality in physics arguably originates with the ``wave-particle'' duality discussed by physicists in what has come to be called the ``old quantum theory'' in the early part of the twentieth century.  (Of course, one can also trace the concept back still further.)  In that context, the idea was that microphysical systems---those systems described by quantum theory---admit of descriptions as constituted by particles and as constituted by waves.  These two descriptions corresponded to distinct conceptual frameworks for understanding the quantum world: on their face, they were incompatible, and yet, there was a sense in which one could, by a change of perspective, reconceive a system in either of these two ways.  This notion of duality was also related to Bohr's ideas about complementarity, according to which there were certain properties of a physical systems---position and momentum, say---that could not be simultaneously ascribed \citep{Bokulich}.  One could conceive of a system as having a definite position, or as having a definite momentum, but not both.

More recent examples of dualities have had a different character \citep{PolchinskiStudies}.  In these cases, one has two theories, each of which can be expressed separately, but which stand in some relationship to one another.  This is to be contrasted with the wave-particle cases, where one had different ``pictures'' of the world, the apparent incompatibility of which was ultimately resolved by moving to a new theory---modern quantum mechanics---that had some of the qualitative features associated with each of the two pictures.  One did not, however, have a well-defined ``wave theory'' and a well-defined ``particle theory'' that were then discovered to stand in some formal relationship to one another.  (This situation regarding wave-particle duality should not be conflated with the equivalence of wave mechanics and matrix mechanics, as shown by \citet{Dirac} and \citet{vonNeumann} and discussed in section \ref{sec:I}; that example is much closer to the sort of duality relationships considered in contemporary physics.)

For example, consider the famous AdS/CFT correspondence \citep{Maldacena,deHaroConceptual,deHaroDEG}, also known as gauge-gravity duality.  According to this duality, a particular (quantum) theory of gravity, satisfying certain geometrical constraints (namely, in which spacetime is asymptotically anti-de Sitter---hence ``AdS''), bears a formal relationship to a (conformal) quantum field theory (the CFT) described on the boundary of the spacetime.  (We note: although many physicists take this duality to be well-established, it is not a mathematical theorem, and indeed, it is not entirely clear how to give a precise mathematical statement of the duality.)  In this case, one has a theory of gravity in $n$ dimensions that is taken to be ``dual'' to a theory without gravity in $n-1$ dimensions.  The sense of ``duality'' here is given by a certain translation manual that allows one to take assertions about states and quantities in one theory and associate them with assertions about states and quantities in the other theory, and vice versa, in a way that bears a passing resemblance to definitional equivalence.

Other examples of dualities studied in string theory include T-duality, in which a theory characterized in a certain spatially compact spacetime with radius $R$ is shown to be dual to another theory in a spatially compact spacetime with radius $1/R$ \citep{tduality1,tduality2}; and S-duality, in which one relates a theory with certain coupling constants $g$ to another theory in which the fields are permuted and the coupling constant becomes $1/g$ \citep{Montonen+Olive,Seiberg+Witten,sduality}.  The term ``duality'' is also used by physicists to describe relationships in, for instance, statistical physics, such as the Kramers-Wannier duality, which relates the free energy of Ising models at different temperatures \citep{KW1,KW2,Wegner}.  These cases, like gauge-gravity duality, involve translation manuals for associating states and quantities between the different theories---though it should be noted that the notion of ``translation'' here does not generally involve precise ``definitions'', as in, for instance, definitional equivalence (recall section \ref{sec:DE}).

Dualities raise a number of questions that have been of interest to philosophers.  Following the physicists who seem to take dual theories to be equivalent descriptions of the world, \citet{deHaroTehButterfield1,deHaroTehButterfield2}, \citet{Rickles}, and \citet{Dawid} have explored the interpretational consequences of such equivalences.  \citet{Rickles}, for instance, argues that dual theories provide apparently different descriptions of the same physical ``structure'', and suggests that any apparent incompatibilities between them should be viewed as illusory or physically insignificant, in the same sense that ``gauge structure'' may be taken to lack physical significance.  He conjectures that given any pair of dual theories, one will always be able to find some third theory that captures the shared structural relationships, without exhibiting any of the apparent inconsistencies.  \citet{deHaroTehButterfield1,deHaroTehButterfield2}, meanwhile, have emphasized the similarities and dissimilarities between dualities and ``gauge'' structure.

On these and similar readings, dualities are examples of ``equivalent theories'' in the wild; they might even be considered as the sorts of test cases that could be used to adjudicate whether various notions of theoretical equivalence discussed by philosophers, such as those described above, are adequate. But this suggestion raises another question.  The examples already mentioned form a diverse bestiary, and there are many other dualities described in the literature \citep[eg. ][]{Witten,Karch+Tong}.  In some cases, these dualities are precisely defined mathematical relationships; in other cases, they are conjectured relationships; and in still other cases, they have a yet more uncertain status.  Given this situation, it is not clear that ``duality'' captures a unique or specific mathematical (or conceptual) relationship.  Could it be that ``duality'' captures a range of different ways in which theories could be related?

In this vein, \citet{deHaro} has proposed a ``schema'' for understanding duality, according to which a duality is an isomorphism between two different possible realizations or formulations of a single physical theory \citep[see also][]{deHaroTehButterfield1,deHaroTehButterfield2,deHaroButterfield}.  (De Haro and collaborators use the term ``model'' to refer to these realizations, but we avoid their usage here, since ``model'' has already been used in a different sense above, and we wish to avoid confusion.  We emphasize that a model in De Haro's sense is more like a ``formulation'' in the sense used above.)  It is essential to this approach that one considers what De Haro calls a ``bare'' theory (or, a ``bare realization'' of that theory), which is a purely formal structure without physical interpretation (or, sometimes, with only partial physical interpretation).  The reason is that dualities, he argues, do not necessarily preserve interpretation: that is, they may relate realizations of a theory that would generally be taken to make different assertions about the world.

More recently, \citet{Butterfield}, building on \citet{deHaroButterfield}, has suggested that De Haro's schema, when applied in some cases, reveals that dualities are \emph{not} necessarily examples of equivalent theories, at least in the sense that philosophers usually have in mind.  The reason is precisely that dualities may not preserve interpretation: de Haro's schema suggests that dualities may exist between theories that are formally equivalent in some sense (in particular, logically equivalent), but which nonetheless make incompatible claims about the world after all.  Observe that this claim is not simply the observation that dual theories may have prima facie different interpretations.  It also involves the claim that, once we \emph{translate} between the theories using whatever manual establishes the existence of the duality, we do not preserve physical meaning.  For instance, in the context of T-duality, one should not say that a theory in which space has radius $R$ is equivalent to one in which space has radius $1/R$, because the quantities $R$ and $1/R$ refer to different radii.  This view amounts to rejecting Rickles' suggestion that dual theories reflect the same underlying physical structure.

\section{Conclusion}

I have discussed several senses in which two physical theories might be said to be equivalent, including that they make the same empirical predictions; they are ``inter-translatable''; they give rise to equivalent categories of models; and they have the same interpretation.  I have also briefly drawn connections between the theoretical equivalence literature in philosophy of science and the literature on dualities in physics.  But many open questions remain.  From my perspective, the most pressing issues concern (a) understanding the character of the gap between generalized definitional / Morita equivalence and categorical equivalence; (b) understanding in what ways and under what circumstances a category of model adequately captures the structure of a physical theory; and (c) further clarifying the relationships between different senses of ``duality'' as they arise in the physics literature, and senses of equivalence as they arise in philosophy of science.

\section*{Acknowledgments}
I am grateful to Thomas Barrett, Clara Bradley, Jeremy Butterfield, Sebastian De Haro, David Malament, James Nguyen, and Nic Teh for comments and suggestions on previous drafts of this article.

\bibliographystyle{elsarticle-harv}
\bibliography{equivalence}

\begin{thebibliography}{127}
\expandafter\ifx\csname natexlab\endcsname\relax\def\natexlab#1{#1}\fi
\expandafter\ifx\csname url\endcsname\relax
  \def\url#1{\texttt{#1}}\fi
\expandafter\ifx\csname urlprefix\endcsname\relax\def\urlprefix{URL }\fi

\bibitem[{Alvarez et~al.(1995)Alvarez, Alvarez-Gaume, and Lozano}]{tduality2}
Alvarez, E., Alvarez-Gaume, L., Lozano, Y., 1995. An introduction to t-duality
  in string theory. Nuclear Physics B-Proceedings Supplements 41~(1-3), 1--20.

\bibitem[{Alvarez-Gaum{\'e} and Hassan(1997)}]{sduality}
Alvarez-Gaum{\'e}, L., Hassan, S., 1997. Introduction to s-duality in n= 2
  supersymmetric gauge theories (a pedagogical review of the work of seiberg
  and witten). Fortschritte der Physik 45~(3-4), 159--236.

\bibitem[{Andr{\'e}ka et~al.(2002)Andr{\'e}ka, Madar{\'a}sz, and
  N{\'e}meti}]{andreka+etal2002}
Andr{\'e}ka, H., Madar{\'a}sz, J.~X., N{\'e}meti, I., 2002. On the logical
  structure of relativity theories. Tech. rep., Alfred R\'enyi Institute of
  Mathematics, Hungarian Academy of Science,
  https://old.renyi.hu/pub/algebraic-logic/Contents.html.

\bibitem[{Andr{\'e}ka and N{\'e}meti(2014)}]{Andreka+NemetiComparing}
Andr{\'e}ka, H., N{\'e}meti, I., 2014. Comparing theories: the dynamics of
  changing vocabulary. In: Johan van Benthem on logic and information dynamics.
  Springer, pp. 143--172.

\bibitem[{Artigue et~al.(1978)Artigue, Isambert, Perrin, and
  Zalc}]{Artigue+etal}
Artigue, M., Isambert, E., Perrin, M., Zalc, A., 1978. Some remarks on
  bicommutability. Fundamenta Mathematicae 101~(3), 207--226.

\bibitem[{Awodey(2006)}]{Awodey}
Awodey, S., 2006. Category Theory. Oxford University Press, New York.

\bibitem[{Awodey and Forssell(2013)}]{AwodeyForssell}
Awodey, S., Forssell, H., 2013. First-order logical duality. Annals of Pure and
  Applied Logic 164~(3), 319--348.

\bibitem[{Baez et~al.(2004)Baez, Bartel, and Dolan}]{Baez+etal}
Baez, J., Bartel, T., Dolan, J., 2004. Property, structure, and stuff,
  available at: http://math.ucr.edu/home/baez/qg-spring2004/discussion.html.

\bibitem[{Barrett(2003)}]{BarrettTheories}
Barrett, J.~A., 2003. Are our best physical theories (probably and/or
  approximately) true? Philosophy of Science 70~(5), 1206--1218.

\bibitem[{Barrett(2014)}]{BarrettCM1}
Barrett, T., 2014. On the structure of classical mechanics. The British Journal
  for the Philosophy of Science 66~(4), 801--828.

\bibitem[{Barrett(2015)}]{BarrettSTS}
Barrett, T.~W., 2015. Spacetime structure. Studies in History and Philosophy of
  Science Part B: Studies in History and Philosophy of Modern Physics 51,
  37--43.

\bibitem[{Barrett(2017{\natexlab{a}})}]{BarrettCM2}
Barrett, T.~W., 2017{\natexlab{a}}. Equivalent and inequivalent formulations of
  classical mechanics. British Journal for Philosophy of ScienceForthcoming.
  http://philsci-archive.pitt.edu/13092/.

\bibitem[{Barrett(2017{\natexlab{b}})}]{BarrettThesis}
Barrett, T.~W., 2017{\natexlab{b}}. On the structure and equivalence of
  theories. Ph.D. thesis, Princeton University,
  http://arks.princeton.edu/ark:/88435/dsp0112579v91j.

\bibitem[{Barrett(2017{\natexlab{c}})}]{barrett2017symmetries}
Barrett, T.~W., 2017{\natexlab{c}}. What do symmetries tell us about
  structure?Forthcoming in Philosophy of Science.

\bibitem[{Barrett and Halvorson(2016{\natexlab{a}})}]{Barrett+HalvorsonGQ}
Barrett, T.~W., Halvorson, H., 2016{\natexlab{a}}. Glymour and {Q}uine on
  theoretical equivalence. Journal of Philosophical Logic 45~(5), 467--483.

\bibitem[{Barrett and Halvorson(2016{\natexlab{b}})}]{Barrett+HalvorsonME}
Barrett, T.~W., Halvorson, H., 2016{\natexlab{b}}. Morita equivalence. The
  Review of Symbolic Logic 9~(3), 556--582.

\bibitem[{Barrett and Halvorson(2017)}]{barrett2017geometry}
Barrett, T.~W., Halvorson, H., 2017. From geometry to conceptual relativity.
  Erkenntnis 82~(5), 1043--1063.

\bibitem[{Bokulich(2017)}]{Bokulich}
Bokulich, P., 2017. Complementarity, wave-particle duality, and domains of
  applicability. Studies in History and Philosophy of Science Part B: Studies
  in History and Philosophy of Modern Physics 59, 136--142.

\bibitem[{Bradley(2018)}]{Bradley}
Bradley, C., 2018. The non-equivalence of einstein and lorentz, unpublished ms.

\bibitem[{Bradley et~al.(2019)Bradley, Rosenstock, and
  Weatherall}]{Bradley+etal}
Bradley, C., Rosenstock, S., Weatherall, J.~O., 2019. Structure, stuff, and
  gauge, unpublished ms.

\bibitem[{Bridgman(1927)}]{Bridgman}
Bridgman, P.~W., 1927. The Logic of Modern Physics. Macmillan, New York, NY.

\bibitem[{Butterfield(2019)}]{Butterfield}
Butterfield, J., 2019. On dualities and equivalences between physical theories.
  In: Huggett, N., W\"uthrich, C. (Eds.), Spacetime After Quantum Gravity.
  Forthcoming.

\bibitem[{Button and Walsh(2018)}]{Walsh+Button}
Button, T., Walsh, S., 2018. Philosophy and model theory. Oxford University
  Press, Oxford.

\bibitem[{Carnap(1958)}]{Carnap}
Carnap, b.~R., 1958. Beobachtungssprache und theoretische sprache. Dialectica
  12~(3-4), 236--248.

\bibitem[{Carnap(1966)}]{CarnapBook}
Carnap, R., 1966. Philosophical Foundations of Physics: An Introduction to the
  Philosophy of Science. Basic Books, New York.

\bibitem[{Coffey(2014)}]{Coffey}
Coffey, K., 2014. Theoretical equivalence as interpretive equivalence,
  forthcoming from \emph{The British Journal for the Philosophy of Science}.

\bibitem[{Cruse(2005)}]{Cruse}
Cruse, P., 2005. Ramsey sentences, structural realism and trivial realization.
  Studies in History and Philosophy of Science 36, 557--–576.

\bibitem[{Curiel(2013)}]{Curiel}
Curiel, E., 2013. Classical mechanics is {L}agrangian; it is not {H}amiltonian.
  The British Journal for Philosophy of Science 65~(2), 269--321.

\bibitem[{Cushing(1994)}]{Cushing}
Cushing, J., 1994. Quantum Mechanics, Historical Contingency, and the
  Copenhagen Hegemony. University of Chicago Press, Chicago, IL.

\bibitem[{Dawid(2017)}]{Dawid}
Dawid, R., 2017. String dualities and empirical equivalence. Studies in History
  and Philosophy of Science Part B: Studies in History and Philosophy of Modern
  Physics 59, 21--29.

\bibitem[{de~Bouvere(1965{\natexlab{a}})}]{Bouvere2}
de~Bouvere, K., 1965{\natexlab{a}}. Logical synonymity. Indagationes
  mathematicae 27, 622--629.

\bibitem[{de~Bouvere(1965{\natexlab{b}})}]{Bouvere1}
de~Bouvere, K., 1965{\natexlab{b}}. Synonymous theories. In: Addison, J.~W.,
  Henkin, L., Tarski, A. (Eds.), The theory of models. North-Holland Pub. Co.,
  Amsterdam, pp. 402--406.

\bibitem[{De~Haro(2017)}]{deHaroDEG}
De~Haro, S., 2017. Dualities and emergent gravity: Gauge/gravity duality.
  Studies in History and Philosophy of Science Part B: Studies in History and
  Philosophy of Modern Physics 59, 109--125.

\bibitem[{De~Haro(2019)}]{deHaro}
De~Haro, S., 2019. Spacetime and physical equivalence. In: Huggett, N.,
  W\"uthrich, C. (Eds.), Spacetime After Quantum Gravity. Forthcoming.
  arXiv:1707.06581.

\bibitem[{De~Haro and Butterfield(2017)}]{deHaroButterfield}
De~Haro, S., Butterfield, J., 2017. A schema for duality, illustrated by
  bosonization. arXiv preprint arXiv:1707.06681.

\bibitem[{De~Haro et~al.(2016{\natexlab{a}})De~Haro, Mayerson, and
  Butterfield}]{deHaroConceptual}
De~Haro, S., Mayerson, D.~R., Butterfield, J.~N., 2016{\natexlab{a}}.
  Conceptual aspects of gauge/gravity duality. Foundations of Physics 46~(11),
  1381--1425.

\bibitem[{De~Haro et~al.(2016{\natexlab{b}})De~Haro, Teh, and
  Butterfield}]{deHaroTehButterfield1}
De~Haro, S., Teh, N., Butterfield, J., 2016{\natexlab{b}}. On the relation
  between dualities and gauge symmetries. Philosophy of Science 83~(5),
  1059--1069.

\bibitem[{De~Haro et~al.(2017)De~Haro, Teh, and
  Butterfield}]{deHaroTehButterfield2}
De~Haro, S., Teh, N., Butterfield, J., 2017. Comparing dualities and gauge
  symmetries. Studies in History and Philosophy of Science Part B: Studies in
  History and Philosophy of Modern Physics 59, 68--80.

\bibitem[{Dewar(2018)}]{DewarRamsey}
Dewar, N., 2018. Ramsey equivalence. ErkenntnisForthcoming.

\bibitem[{Dewar and Eva(2017)}]{Dewar+Eva}
Dewar, N., Eva, B., 2017. A categorical perspective on symmetry and
  equivalence.

\bibitem[{Dirac(1930)}]{Dirac}
Dirac, P. A.~M., 1930. The Principles of Quantum Mechanics. Oxford University
  Press, Oxford.

\bibitem[{Downes(1992)}]{Downes}
Downes, S.~M., 1992. The importance of models in theorizing: A deflationary
  semantic view. In: PSA: Proceedings of the biennial meeting of the philosophy
  of science association. Vol. 1992. Philosophy of Science Association, pp.
  142--153.

\bibitem[{D{\"u}rr et~al.(2012)D{\"u}rr, Goldstein, and
  Zangh{\`\i}}]{Durr+etal}
D{\"u}rr, D., Goldstein, S., Zangh{\`\i}, N., 2012. Quantum physics without
  quantum philosophy. Springer Science \& Business Media, Heidelberg.

\bibitem[{Giveon et~al.(1994)Giveon, Porrati, and Rabinovici}]{tduality1}
Giveon, A., Porrati, M., Rabinovici, E., 1994. Target space duality in string
  theory. Physics Reports 244~(2-3), 77--202.

\bibitem[{Glymour(1970)}]{GlymourTETR}
Glymour, C., 1970. Theoretical equivalence and theoretical realism. PSA:
  Proceedings of the Biennial Meeting of the Philosophy of Science Association
  1970, 275--288.

\bibitem[{Glymour(1980)}]{GlymourTE}
Glymour, C., 1980. Theory and Evidence. Princeton University Press, Princeton,
  NJ.

\bibitem[{Glymour(2013)}]{GlymourHalvorson}
Glymour, C., 2013. Theoretical equivalence and the semantic view of theories.
  Philosophy of Science 80~(2), 286--297.

\bibitem[{Gr{\"u}nbaum(1962)}]{Grunbaum}
Gr{\"u}nbaum, A., 1962. Geometry, chronometry and empiricism, 405--526.

\bibitem[{Halvorson(2012)}]{Halvorson}
Halvorson, H., 2012. What scientific theories could not be. Philosophy of
  Science 79~(2), 183--206.

\bibitem[{Halvorson(2013)}]{HalvorsonGlymour}
Halvorson, H., 2013. The semantic view, if plausible, is syntactic. Philosophy
  of Science 80~(3), 475--478.

\bibitem[{Halvorson(2015{\natexlab{a}})}]{HalvorsonCategories}
Halvorson, H., 2015{\natexlab{a}}. Categories of scientific theories. In:
  Landry, E. (Ed.), Categories for the Working Philosopher. Oxford University
  Press, Oxford, UK, this volume.

\bibitem[{Halvorson(2015{\natexlab{b}})}]{HalvorsonOxford}
Halvorson, H., 2015{\natexlab{b}}. Scientific theories. In: Humphreys, P.
  (Ed.), The Oxford Handbook of the Philosophy of Science. Oxford University
  Press, Oxford, UK, . Forthcoming. http://philsci-archive.pitt.edu/11347/.

\bibitem[{Halvorson and Tsementzis(2017)}]{HalvorsonLandryVolume}
Halvorson, H., Tsementzis, D., 2017. Categories of scientific theories. In:
  Landry, E. (Ed.), Categories for the working philosopher. Oxford University
  Press, Oxford, pp. 402--429.

\bibitem[{Hempel(1958)}]{Hempel1}
Hempel, C.~G., 1958. The theoretician's dilemma: A study in the logic of theory
  construction. In: Feigl, H., Scriven, M., Maxwell, G. (Eds.), Concepts,
  Theories, and the Mind-Body Problem. University of Minnesota Press,
  Minneapolis, MN, pp. 37--98.

\bibitem[{Hempel(1973)}]{Hempel2}
Hempel, C.~G., 1973. The meaning of theoretical terms: A critique of the
  standard empiricist construal. In: Suppes, P., Henkin, L., Joja, A., Moisil,
  G.~C. (Eds.), Logic, Methodology and Philosophy of Science IV. North-Holland
  Publishing Co., Amsterdam, pp. 367--–378.

\bibitem[{Hertz(1899 [1894])}]{Hertz}
Hertz, H., 1899 [1894]. The Principles of Mechanics Presented in a New Form.
  MacMillan \& Co., London.

\bibitem[{Hodges(1993)}]{Hodges}
Hodges, W., 1993. Model theory. Vol.~42. Cambridge University Press.

\bibitem[{Hodges(1997)}]{hodges1997shorter}
Hodges, W., 1997. A shorter model theory. Cambridge university press,
  Cambridge, UK.

\bibitem[{Hudetz(2017)}]{hudetz2017semantic}
Hudetz, L., 2017. The semantic view of theories and higher-order languages.
  Synthese, 1--19.

\bibitem[{Hudetz(2018)}]{Hudetz}
Hudetz, L., 2018. Definable categorical equivalence. Philosophy of
  ScienceForthcoming. http://philsci-archive.pitt.edu/14297/.

\bibitem[{Kanger(1968)}]{Kanger}
Kanger, S., 1968. Equivalent theories. Theoria 34~(1), 1--6.

\bibitem[{Karch and Tong(2016)}]{Karch+Tong}
Karch, A., Tong, D., 2016. Particle-vortex duality from 3d bosonization.
  Physical Review X 6~(3), 031043.

\bibitem[{Ketland(2004)}]{Ketland}
Ketland, J., 2004. Empirical adequacy and ramsification. The British Journal
  for the Philosophy of Science 55~(2), 287--300.

\bibitem[{Kramers and Wannier(1941{\natexlab{a}})}]{KW1}
Kramers, H.~A., Wannier, G.~H., 1941{\natexlab{a}}. Statistics of the
  two-dimensional ferromagnet. part i. Physical Review 60~(3), 252.

\bibitem[{Kramers and Wannier(1941{\natexlab{b}})}]{KW2}
Kramers, H.~A., Wannier, G.~H., 1941{\natexlab{b}}. Statistics of the
  two-dimensional ferromagnet. part ii. Physical Review 60~(3), 263.

\bibitem[{{Le Bihan} and Read(2019)}]{LeBihan+Read}
{Le Bihan}, B., Read, J., 2019. Duality and ontology. Philosophy
  CompassForthcoming.

\bibitem[{Lefever and Sz{\'e}kely(2017)}]{Lefever+SzekelyComparing}
Lefever, K., Sz{\'e}kely, G., 2017. Comparing classical and relativistic
  kinematics in first-order logic, arXiv:1707.05371.

\bibitem[{Lefever and Sz{\'e}kely(2018)}]{Lefever+Szekely}
Lefever, K., Sz{\'e}kely, G., 2018. On generalization of definitional
  equivalence to languages with non-disjoint signatures, arXiv:1802.06844.

\bibitem[{Leinster(2014)}]{Leinster}
Leinster, T., 2014. Basic Category Theory. Cambridge University Press,
  Cambridge.

\bibitem[{Lewis(1970)}]{Lewis}
Lewis, D., 1970. How to define theoretical terms. Journal of Philosophy
  67~(13), 427–--446.

\bibitem[{Lewis(1972)}]{lewis1972psychophysical}
Lewis, D., 1972. Psychophysical and theoretical identifications. Australasian
  Journal of Philosophy 50~(3), 249--258.

\bibitem[{Lutz(2014)}]{lutz2014s}
Lutz, S., 2014. What’s right with a syntactic approach to theories and
  models? Erkenntnis 79~(8), 1475--1492.

\bibitem[{Lutz(2017)}]{lutz2017syntax}
Lutz, S., 2017. What was the syntax-semantics debate in the philosophy of
  science about? Philosophy and Phenomenological Research 95~(2), 319--352.

\bibitem[{{Mac Lane}(1998)}]{MacLane}
{Mac Lane}, S., 1998. Categories for the Working Mathematician, 2nd Edition.
  Springer, New York.

\bibitem[{Madar{\'a}sz(2002)}]{Madarasz}
Madar{\'a}sz, J.~X., 2002. Logic and relativity (in the light of definability
  theory). Ph.D. thesis, Eotv\"os Lor\"and University, Budapest.

\bibitem[{Magnus and Frost-Arnold(2010)}]{FrostArnold+Magnus}
Magnus, P., Frost-Arnold, G., 2010. The identical rivals response to
  underdetermination. In: Magnus, P., Busch, J. (Eds.), New Waves in Philosophy
  of Science. Palgrave Macmillan, London, pp. 112--130.

\bibitem[{Makkai(1993)}]{Makkai}
Makkai, M., 1993. Duality and definability in first order logic. Vol. 503.
  American Mathematical Soc., Providence, RI.

\bibitem[{Malament(1995)}]{MalamentNorton}
Malament, D., 1995. Is {N}ewtonian cosmology really inconsistent? Philosophy of
  Science 62~(4), 489--510.

\bibitem[{Malament(2012)}]{MalamentGR}
Malament, D., 2012. Topics in the Foundations of General Relativity and
  {N}ewtonian Gravitation Theory. University of Chicago Press, Chicago.

\bibitem[{Maldacena(1999)}]{Maldacena}
Maldacena, J., 1999. The large-n limit of superconformal field theories and
  supergravity. International journal of theoretical physics 38~(4),
  1113--1133.

\bibitem[{Matsubara(2013)}]{Matsubara}
Matsubara, K., 2013. Realism, underdetermination and string theory dualities.
  Synthese 190~(3), 471--489.

\bibitem[{Maudlin(2018)}]{Maudlin}
Maudlin, T., 2018. Ontological clarity via canonical presentation:
  Electromagnetism and the aharonov--bohm effect. Entropy 20~(6), 465.

\bibitem[{Maxwell(1962)}]{MaxwellG1}
Maxwell, G., 1962. The ontological status of theoretical entities. In: Feigl,
  H., Maxwell, G. (Eds.), Scientific Explanation, Space, and Time. University
  of Minnesota Press, Minneapolis, MN, pp. 3--14.

\bibitem[{Maxwell(1970)}]{MaxwellG2}
Maxwell, G., 1970. Structural realism and the meaning of theoretical terms. In:
  Winokur, S., Radner, M. (Eds.), Analyses of Theories and Methods of Physics
  and Psychology. University of Minnesota Press, Minneapolis, pp. 181--192.

\bibitem[{Melia and Saatsi(2006)}]{Melia+Saatsi}
Melia, J., Saatsi, J., 2006. Ramseyfication and theoretical content. The
  British Journal for the Philosophy of Science 57~(3), 561--585.

\bibitem[{Montague(1957)}]{Montague}
Montague, R., 1957. Contributions to the axiomatic foundations of set theory.
  Ph.D. thesis, University of California, Berkeley.

\bibitem[{Montonen and Olive(1977)}]{Montonen+Olive}
Montonen, C., Olive, D., 1977. Magnetic monopoles as gauge particles? Physics
  Letters B 72~(1), 117--120.

\bibitem[{Muller(1997{\natexlab{a}})}]{MullerMyth1}
Muller, F.~A., 1997{\natexlab{a}}. The equivalence myth of quantum
  mechanics—part i. Studies in History and Philosophy of Science Part B:
  Studies in History and Philosophy of Modern Physics 28~(1), 35--61.

\bibitem[{Muller(1997{\natexlab{b}})}]{MullerMyth2}
Muller, F.~A., 1997{\natexlab{b}}. The equivalence myth of quantum
  mechanics—part ii. Studies in History and Philosophy of Science Part B:
  Studies in History and Philosophy of Modern Physics 28~(2), 219--247.

\bibitem[{Nguyen(2017)}]{NguyenTE}
Nguyen, J., 2017. Scientific representation and theoretical equivalence.
  Philosophy of Science 84~(5), 982--995.

\bibitem[{Nguyen et~al.(2018)Nguyen, Teh, and Wells}]{Nguyen+etal}
Nguyen, J., Teh, N.~J., Wells, L., 2018. Why surplus structure is not
  superfluous. British Journal for Philosophy of ScienceForthcoming.

\bibitem[{North(2009)}]{North}
North, J., 2009. The `structure' of physics: A case study. Journal of
  Philosophy 106~(2), 57--88.

\bibitem[{Norton(1992)}]{NortonCosmology}
Norton, J., 1992. A paradox in {N}ewtonian gravitation theory. PSA: Proceedings
  of the Biennial Meeting of the Philosophy of Science Association 1992,
  412--420.

\bibitem[{Norton(1995)}]{NortonAcceleration}
Norton, J., 1995. The force of {N}ewtonian cosmology: Acceleration is relative.
  Philosophy of Science 62~(4), 511--522.

\bibitem[{Norton(2008)}]{NortonUnderdetermination}
Norton, J., 2008. Must evidence underdetermine theory. In: Kourany, J.~A.,
  Carrier, M., Howard, D. (Eds.), The challenge of the social and the pressure
  of practice: Science and values revisited. University of Pittsburgh Press
  Pittsburgh, Pittsburgh, PA, pp. 17--44.

\bibitem[{O'Connor and Weatherall(2016)}]{OConnor+Weatherall}
O'Connor, C., Weatherall, J.~O., 2016. Black holes, black-scholes, and prairie
  voles: An essay review of simulation and similarity, by michael weisberg.
  Philosophy of Science 83~(4), 613--626.

\bibitem[{Polchinski(2017)}]{PolchinskiStudies}
Polchinski, J., 2017. Dualities of fields and strings. Studies in History and
  Philosophy of Science Part B: Studies in History and Philosophy of Modern
  Physics 59, 6--20.

\bibitem[{Psillos(2000)}]{Psillos}
Psillos, S., 2000. Carnap, the ramsey-sentence and realistic empiricism.
  Erkenntnis 52~(2), 253--279.

\bibitem[{Quine(1975)}]{Quine}
Quine, W.~V., 1975. On empirically equivalent systems of the world. Erkenntnis
  9~(3), 313--328.

\bibitem[{Ramsey(1931)}]{Ramsey}
Ramsey, F.~P., 1931. The Foundations of Mathematics. Routledge \& Kegan Paul,
  London, UK, Ch. Theories, pp. 212--236.

\bibitem[{Read(2016)}]{Read}
Read, J., 2016. The interpretation of string-theoretic dualities. Foundations
  of Physics 46~(2), 209--235.

\bibitem[{Read and M\/oller-Nielsen(2018)}]{Read+MN}
Read, J., M\/oller-Nielsen, T., 2018. Motivating dualities.
  SyntheseForthcoming.

\bibitem[{Reichenbach(1938)}]{Reichenbach}
Reichenbach, H., 1938. Experience and prediction: An analysis of the
  foundations and the structure of knowledge. University of Chicago Press,
  Chicago, IL.

\bibitem[{Rickles(2011)}]{RicklesOld}
Rickles, D., 2011. A philosopher looks at string dualities. Studies in Histoy
  and Philosophy of Modern Physics 42~(1), 54--67.

\bibitem[{Rickles(2017)}]{Rickles}
Rickles, D., 2017. Dual theories: `same but different' or `different but same'?
  Studies in History and Philosophy of Science Part B: Studies in History and
  Philosophy of Modern Physics 59, 62--67.

\bibitem[{Rosenstock et~al.(2015)Rosenstock, Barrett, and
  Weatherall}]{Rosenstock+etal}
Rosenstock, S., Barrett, T.~W., Weatherall, J.~O., 2015. On {E}instein algebras
  and relativistic spacetimes. Studies in History and Philosophy of Science
  Part B: Studies in History and Philosophy of Modern Physics 52, 309--316.

\bibitem[{Rosenstock and Weatherall(2016)}]{Rosenstock+Weatherall}
Rosenstock, S., Weatherall, J.~O., 2016. A categorical equivalence between
  generalized holonomy maps on a connected manifold and principal connections
  on bundles over that manifold. Journal of Mathematical Physics 57~(10),
  102902.

\bibitem[{Rynasiewicz(1992)}]{Ryno}
Rynasiewicz, R., 1992. Rings, holes and substantivalism: On the program of
  leibniz algebras. Philosophy of Science 59~(4), 572--589.

\bibitem[{Salmon(1966)}]{Salmon}
Salmon, W.~C., 1966. Verifiability and logic. In: Feyerabend, P.~K., Maxwell,
  G. (Eds.), Mind, matter and method: essays in philosophy and science in honor
  of Herbert Feigl. University of Minnesota Press, Minneapolis, MN, pp.
  354--366.

\bibitem[{Schreiber(2013)}]{Schreiber}
Schreiber, U., 2013. Differential cohomology in a cohesive infinity-topos,
  arXiv:1310.7930.

\bibitem[{Schreiber and Waldorf(2007)}]{Schreiber+Waldorff}
Schreiber, U., Waldorf, K., 2007. Parallel transport and
  functorsArXiv:0705.0452.

\bibitem[{Seiberg and Witten(1994)}]{Seiberg+Witten}
Seiberg, N., Witten, E., 1994. Electric-magnetic duality, monopole
  condensation, and confinement in n= 2 supersymmetric yang-mills theory.
  Nuclear Physics B 426~(1), 19--52.

\bibitem[{Sklar(1982)}]{Sklar}
Sklar, L., 1982. Saving the noumena. Philosophical Topics 13~(1), 89--110.

\bibitem[{Trautman(1965)}]{Trautman}
Trautman, A., 1965. Foundations and current problem of general relativity. In:
  Deser, S., Ford, K.~W. (Eds.), Lectures on General Relativity. Prentice-Hall,
  Englewood Cliffs, NJ, pp. 1--248.

\bibitem[{{Univalent Foundations Program}(2013)}]{hottbook}
{Univalent Foundations Program}, T., 2013. Homotopy Type Theory: Univalent
  Foundations of Mathematics. \url{https://homotopytypetheory.org/book},
  Institute for Advanced Study.

\bibitem[{Van~Fraassen(2014)}]{BvF}
Van~Fraassen, B.~C., 2014. One or two gentle remarks about hans halvorson’s
  critique of the semantic view. Philosophy of Science 81~(2), 276--283.

\bibitem[{Visser(2017)}]{Visser}
Visser, A., 2017. Categories of theories and interpretations. In: Enayat, A.,
  Kalantari, I., Moniri, M. (Eds.), Logic in Tehran. Cambridge University
  Press, Cambridge, UK, pp. 284--341.

\bibitem[{Von~Neumann(1955 [1932])}]{vonNeumann}
Von~Neumann, J., 1955 [1932]. Mathematical foundations of quantum theory.
  Princeton University Press.

\bibitem[{Wallace(2017)}]{WallaceNewtCosm}
Wallace, D., 2017. More problems for {N}ewtonian cosmology. Studies in History
  and Philosophy of Science Part B: Studies in History and Philosophy of Modern
  Physics 57, 35--40.

\bibitem[{Weatherall(2016{\natexlab{a}})}]{WeatherallTE}
Weatherall, J.~O., 2016{\natexlab{a}}. Are {N}ewtonian gravitation and
  geometrized {N}ewtonian gravitation theoretically equivalent? Erkenntnis
  81~(5), 1073--1091.

\bibitem[{Weatherall(2016{\natexlab{b}})}]{WeatherallUG}
Weatherall, J.~O., 2016{\natexlab{b}}. Understanding gauge. Philosophy of
  Science 83~(5), 1039--1049.

\bibitem[{Weatherall(2017)}]{WeatherallLandry}
Weatherall, J.~O., 2017. Category theory and the foundations of classical
  space-time theories. In: Landry, E. (Ed.), Categories for the Working
  Philosopher. Oxford University Press, Oxford, pp. 329--348.

\bibitem[{Weatherall(2018)}]{WeatherallWNCE}
Weatherall, J.~O., 2018. Why not categorical equivalence?, unpublished ms.

\bibitem[{Wegner(1971)}]{Wegner}
Wegner, F.~J., 1971. Duality in generalized {I}sing models and phase
  transitions without local order parameters. Journal of Mathematical Physics
  12~(10), 2259--2272.

\bibitem[{Weisberg(2012)}]{Weisberg}
Weisberg, M., 2012. Simulation and similarity: Using models to understand the
  world. Oxford University Press, New York.

\bibitem[{Witten(1994)}]{Witten}
Witten, E., 1994. Non-{A}belian bosonization in two dimensions. In:
  Bosonization. World Scientific, pp. 201--218.

\bibitem[{Worrall(2007)}]{Worrall}
Worrall, J., 2007. Miracles and models: Why reports of the death of structural
  realism may be exaggerated. Royal Institute of Philosophy Supplements
  82~(61), 125--–154.

\end{thebibliography}

\end{document}